\documentclass[12pt]{article}
\usepackage{amsmath}
\usepackage{amssymb}
\usepackage{epsf}

\newfont{\bit}{cmbxti10 scaled 1728}
\renewcommand{\thefootnote}{\fnsymbol{footnote}}

\begin{document}
\newpage
\pagestyle{empty}

\begin{center}
{\LARGE {Distributional energy-momentum tensor\\
 of\\
the extended Kerr geometry\\ 
}}

\vspace{2cm}
{\large Herbert BALASIN
\footnote[1]{e-mail: hbalasin@unix.uvic.ca}
\footnote[4]{supported by the APART-program of the 
Austrian Academy of Sciences}
}\\[.3cm]
{\em
Department of Physics and Astronomy, University of Victoria\\
 Victoria B.C.~, V8W 3P6, CANADA
}\\[.8cm]
\begin{minipage}[t]{11.5cm}
\hspace{-1cm}{\bf Dedication:} I would like to dedicate this 
article to my friend Michaela Hraby who was killed in a car accident for 
all her support and inspiration.
\end{minipage}
\end{center}
\vspace{1.5cm}
\begin{abstract}
We generalize previous work on the energy-momentum tensor-distribution
of the Kerr geometry by extending the manifold structure into the
negative mass region. Since the extension of the flat part of the 
Kerr-Schild decomposition from one sheet to the double cover
develops a singularity at the branch surface we have to take
its non-smoothness into account. It is however possible to find a 
geometry within the generalized Kerr-Schild class that is in the 
Colombeau-sense associated to the maximally analytic Kerr-metric.

\end{abstract}
\vfill

\rightline{TUW 97 -- 03}
\rightline{January 1997}

\renewcommand{\thefootnote}{\arabic{footnote}}
\setcounter{footnote}{0}
\newpage  
\pagebreak
\pagenumbering{arabic}
\pagestyle{plain}
\renewcommand{\baselinestretch}{1}
\small\normalsize 
\section*{\large\bit Introduction}
Although the Kerr geometry is usually considered to be vacuum solution
of the Einstein equations, there has been quite some interest in the
investigation of its (singular) source structure 
\cite{Isr1,Isr2,Lop,Bur,BaNa2}.
Aside from shedding light on the singularity structure the knowledge of the 
source, specifically the energy-momentum tensor, provides a useful tool
in the calculation of the ultrarelativistic limit geometries of the 
Kerr spacetime \cite{BaNa3,BaNa4}. This is mainly due to the fact that the 
ambiguities encountered in the metric-limit \cite{LouSa} are not present 
in the approach based on the energy-momentum tensor \cite{BaNa3}. \par
In a previous paper \cite{BaNa2} we made use of the Kerr-Schild decomposition
\cite{KeSch}
to assign an energy-momentum tensor to the Kerr-Newman 
spacetime-family. The important observation in this context was the fact
that the mixed form of the Ricci tensor (and thereby the 
energy-momentum tensor) is linear in the profile $\hat{f}$ of the 
decomposition. This allows an unambiguous distributional treatment 
despite the nonlinear character of Einstein's theory of gravity. 
As expected the energy-momentum tensor-distribution has its support 
on the singular region of the geometry, which in the case of Kerr is a ring. 
However, since we restricted the manifold to the region of positive mass
(or positive radial coordinate $r$) we picked up an additional
contribution which is concentrated on the on the disk spanned by the
ring. This contribution is due to the fact that we identified the
two different boundaries of the branch cut. The aim of the
present paper is to extend our treatment to the maximal analytic
extension, which includes both sheets and thereby avoids the
previous identification. The flat metric $\eta_{ab}$ in the 
Kerr-Schild decomposition develops a singularity along the branch 
surface, which does not allow the immediate application of the
Kerr-Schild method we used in \cite{BaNa2}. Within the
generalized Kerr-Schild class \cite{Taub} it is however possible to
find a smooth metric that is Colombeau-associated \cite{Col1} 
to the Kerr-solution. 
Calculating its energy-momentum tensor allows 
to perform the distributional limit in the end.\par
Our work is organized in the following way: 
In section one we review the generalized Kerr-Schild class
and present a conceptive example displaying the curvature of the 
quasi-flat part of decomposition when extended to the double-cover. 
In section two we briefly recapitulate the basic concepts of Colombeau's
algebra of generalized functions. Finally section three contains 
the regularisation of the Kerr geometry within the
generalized Kerr-Schild class and the calculation of its
energy-momentum tensor.

\section*{\large\bit 1) Generalized Kerr-Schild class}

Usually geometries in the Kerr-Schild class are defined by 
decomposing the metric

\begin{equation}\label{ksmetric}
  g_{ab} = \eta_{ab} + \hat{f}\> k_a k_b\qquad k_a := \eta_{ab}k^b\quad
  k^a k^b\eta_{ab}=0,
\end{equation}
where $\eta_{ab}$ denotes the flat part of the decomposition and
$k^a$ is a null vector-field with respect to both geometries.
An immediate consequence of (\ref{ksmetric}) is 
$$
(k\nabla)k^a = (k\partial)k^a,
$$
where $\nabla_a$ and $\partial_a$ are the covariant derivatives of 
$g_{ab}$ and $\eta_{ab}$ respectively. In particular requiring $k^a$
to be geodetic with respect to $\eta_{ab}$ entails its geodeticity with
respect to $g_{ab}$. Geometries with the above properties define
the Kerr-Schild class. Its most important representative is the
three-parameter Kerr-Newman family of (electro-)vacuum geometries,
whose most general element describes a stationary, rotating, charged
black hole. A simple subclass consists of geometries where the covariant
constancy of $k^a$ is required. These are the so-called plane-waves 
with parallel rays (pp-waves) \cite{JEK}. \par
An important property of the Kerr-Schild class is the fact
that the mixed Ricci-tensor is linear in the function $\hat{f}$ (which will
be referred to as the profile by borrowing terminology from the pp-wave case):
\begin{equation}\label{ksricci}
  R^a{}_b = \frac{1}{2}\left[ \partial^a\partial_c(\hat{f}k^c k_b) +
\partial_b\partial_c(\hat{f}k^c k^a) - \partial^2(\hat{f}k^a k_b)\right]
\end{equation}
It is remarkable that all of the above properties remain intact if one
replaces the flat background by an arbitrary one, i.~e.~one considers
geometries of the form
\begin{equation}\label{gksmetric}
  \tilde{g}_{ab} = g_{ab} + \hat{f}\> k_a k_b,\qquad k_a := g_{ab}k^b\quad
  k^a k^b g_{ab}=0.
\end{equation}
Once again geodeticity of $k^a$ with respect to $g_{ab}$ implies
the according behavior with respect to $\tilde{g}_{ab}$ since
$$
(k\tilde{\nabla})k^a = (k\nabla)k^a
$$
holds. The expression for the Ricci tensor (\ref{ksricci}) becomes modified 
\begin{align}\label{gksricci}
  \tilde{R}^a{}_b &= R^a{}_b - 
\frac{1}{2}\left[R^a{}_c\>\hat{f}k^c k_b - R_{bc}\>\hat{f}k^c k^a
- 2R^a{}_{cdb}\>\hat{f}k^ck^d\right.\nonumber\\
 &\left.+ (\nabla^a\nabla_c(\hat{f}k^c k_b) + \nabla_b\nabla_c(\hat{f}k^c k^a) - 
\nabla^2(\hat{f}k^a k_b))\right],
\end{align}
however the linearity in $\hat{f}$ remains.\par
\begin{figure}[htbp]
  \begin{center}
    \leavevmode
\epsfysize=4.5cm
\epsfbox{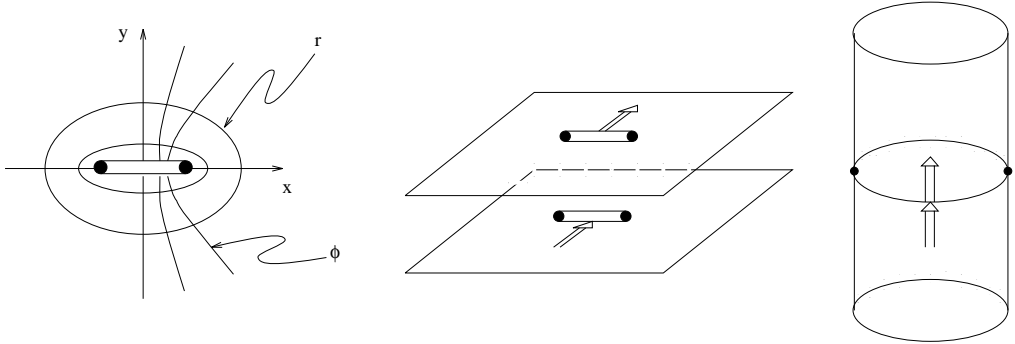}
    \caption{double cover of ${\mathbb R}^2$}
    \label{fig:1}
  \end{center}
\end{figure}

Let us now turn to the Kerr geometry in coordinates adapted to (\ref{ksmetric})
\begin{align}\label{kmetric}
  ds^2 &= -dt^2 + \frac{\Sigma}{r^2 + a^2} dr^2 + \Sigma d\theta^2
+ (r^2 + a^2)\sin^2\theta d\phi^2 + \nonumber\\&+ \frac{2mr}{\Sigma}(dt - 
\frac{\Sigma}{r^2+a^2}dr + a \sin^2\theta d\phi)^2
\qquad\Sigma=r^2 + a^2\cos^2\theta.
\end{align}
The first line of (\ref{kmetric}) is usually considered to represent 
Minkowski space in spheroidal coordinates, however it is possible to exhibit
the double cover structure of the maximal analytic continuation by allowing 
$r$ to take negative values as well. 
This fact may be displayed more transparently if we consider
a two-dimensional analogue of the spatial part of the flat
background. For this end let us consider ${\mathbb R}^2$ in elliptical 
coordinates.
\begin{align}\label{twoex}
  ds^2&=dx^2+dy^2=\frac{\Sigma}{r^2+a^2}dr^2 +\Sigma d\phi^2
  \qquad &&x=\sqrt{r^2+a^2}\cos\phi\nonumber\\
  &\Sigma=r^2+a^2\sin^2\phi &&y=r\sin\phi 
\end{align}
By extending the range of $r$ to negative values, it is
possible to extend (\ref{twoex}) to a cylinder as shown in fig.~\ref{fig:1}

\noindent
The singularities are located at the branch-points $r=0,\phi=0$ and $\pi$, 
which is also the location where the curvature is concentrated.
In order to show this let us ``regularize'' (\ref{twoex}) by replacing
$r^2\rightarrow r^2 +\alpha^2$.
\begin{align}\label{twoexreg}
  ds^2 &= a^2(\frac{f^2}{g^2}\cosh^2u du^2 + f^2 d\phi^2)\qquad
  &&f^2 = \sinh^2 u + \sin^2 \phi + \beta^2\nonumber\\
  &r=a\sinh u,\> \alpha= a \beta &&g^2 = \cosh^2 u + \beta^2
\end{align}
Using an adapted frame one obtains the connection and the Riemann-tensor
\begin{align}
  &e^u = a\frac{f}{g}du\qquad
  &&\omega^u{}_\phi =\frac{\cos\phi\sin\phi\cosh u}{f^2g}du
  - \frac{g\sinh u}{f^2}d\phi\nonumber\\
  &e^\phi = af d\phi
  &&R^u{}_\phi = -\frac{\beta^2(f^2 + 2 \cos\phi)\cosh u}{f^4 g} du d\phi.
\end{align}
Evaluating the densitized Ricci tensor 
\begin{align}
  &\sqrt{g}R^a{}_b = -\frac{\beta^2(f^2+2\cos^2\phi)\cosh u}{f^4g}
  (\partial_u^a du_b + \partial_\phi^a d\phi_b) =: F(x)
  (\partial_u^a du_b + \partial_\phi^a d\phi_b)\nonumber\\
\intertext{on an arbitrary test-function $\varphi(x)$ gives}
  &\lim\limits_{\beta\to 0}\int F(x)\varphi(x)d^2x =
  -\int \frac{\beta^2(f^2+2\cos^2\phi)\cosh u}{f^4g}\varphi(u,\phi)du d\phi
  \nonumber\\
  &=-\left(\int\limits_{-\infty}^{\infty}\hspace{-0.2cm}du\hspace{-0.1cm}
  \int\limits_{-\frac{\pi}{2}}^{\frac{\pi}{2}}\hspace{-0.2cm}d\phi
  \>\varphi(u,\phi) +
  \int\limits_{-\infty}^{\infty}\hspace{-0.2cm}du\hspace{-0.1cm}
  \int\limits_{-\frac{\pi}{2}}^{\frac{\pi}{2}}\hspace{-0.2cm}d\phi
  \>\varphi(u,\phi+\pi)\right)
  \frac{\beta^2(f^2+2\cos^2\phi)\cosh u}{f^4g}\nonumber\\
  &=-2\pi(\varphi(0,0)+\varphi(0,\pi))=-2\pi(\delta(u)\delta(\sin\phi),\varphi)
\end{align}
which explicitly shows that (\ref{twoex}) when extended to the cylinder
is no longer flat but develops curvature concentrated on the branch points.
Since the main difference between (\ref{kmetric}) between and (\ref{twoex})
is the breaking of the $U(1)$ symmetry down to ${\mathbb Z}_2$ 
it is obvious that we can no longer use the former as part of the smooth
structure. 

\section*{\large\bit 2) A short review of Colombeau theory}

The expression for the Ricci tensor of the generalized Kerr-Schild
class involves terms which are products of singular quantities like
the curvature of the background and the profile $\hat{f}$. Within distribution
theory such products are in general meaningless. There exists however
a recent generalization due to Colombeau \cite{Col1,Col2} which embeds 
distribution space $\mathcal D'$ into the larger algebra $\mathcal G$ of 
generalized functions. We will sketch the ideas of its construction only 
briefly and refer the reader to \cite{Col1,Col2,ArBi,Ba2} for a
more detailed treatment. The main idea is to consider one-parameter 
families of $C^\infty$ functions $(f_\epsilon)_{0<\epsilon<1}$ as basic
objects\footnote{Intuitively distributions correspond to the limit
$\epsilon\to 0$ and $f_\epsilon(x)$ represents the additional information lost
in the process of idealization.}.
They form an algebra under the naturally defined pointwise
operations. The usual $C^\infty$ functions are embedded as constant sequences,
which does not require any additional structure. On the other hand
the embedding of $C^p$ functions and distributions is achieved by
convolution with an appropriate smoothing kernel $\rho$ via
\begin{equation*}
  f_\epsilon(x) = \int d^ny\frac{1}{\epsilon^n}\rho(\frac{y-x}{\epsilon})f(y).
\end{equation*}
Since $C^\infty$ functions are obviously of class $C^p$, consistency requires 
the identification of the different embeddings. The difference of the 
two embeddings belongs to the set of so-called negligible families,
which vanish faster than any given positive power of $\epsilon$ on any 
given compact set. Since we want this set to be an ideal, in order to preserve
the algebra structure under the identification, one has to require
a growth condition in $\epsilon$ on the general families. These so-called
moderate families do not grow faster than inverse powers of $\epsilon$ 
in the limit $\epsilon\to 0$. It can be shown that the embedding of 
distributions generates moderate families \cite{Col1,Col2}.\par
Generalized functions (elements of $\mathcal{G}$) are thus equivalence classes
of moderate families modulo negligible ones. (This situation is actually very
similar to the one encountered in $L^p$ spaces where in our case the 
negligible functions play the role of measurable functions that vanish 
outside a set of measure zero.)\par
The contact with usual distribution theory is achieved by coarse-graining 
$\mathcal G$. The idea is to pack together different Colombeau-objects
that give the same distribution in the limit $\epsilon\to 0$:
\begin{align*}
  &(f_\epsilon)\in{\mathcal G} \approx T\in{\mathcal
D}'\quad\text{if}\qquad \forall\varphi\in
{\mathcal D}\qquad\lim_{\epsilon\to 0}\int dx
f_\epsilon(x)\varphi(x)  - (T,\varphi) =0,\nonumber\\ 
\intertext{or more generally}
  &(f_\epsilon)\in{\mathcal G} \approx (g_\epsilon)\in{\mathcal G}
  \quad\text{if}\qquad\forall\varphi\in
{\mathcal D}\qquad\lim_{\epsilon\to 0}
  \int dx(f_\epsilon(x)-g_\epsilon(x))\varphi(x)=0.
\end{align*}
The equivalence relation, which is usually called association, 
respects addition, multiplication by $C^\infty$
functions and differentiation. However it does not respect 
multiplication, which is to be expected since it models distributional
equality within $\mathcal G$. Let us now come back to the question raised at
the beginning of this section which is essentially boils down to the
existence of a distribution that is associated with the product of 
$P(1/x)$ with $\delta(x)$. Within $\cal G$ we have the following 
representations
$$
P_\epsilon(\frac{1}{x})= \frac{x}{x^2+\epsilon^2}
\quad\text{and}\quad 
\delta_\epsilon(x)= \frac{\epsilon}{\pi}\frac{1}{x^2+\epsilon^2}.
$$
In order to find the a distribution associated to their product we have to
evaluate
$$
\lim_{\epsilon\to 0}\int dx \frac{\epsilon x}{\pi(x^2+\epsilon^2)^2}
\varphi(x)=\int dx \frac{1}{\epsilon\pi}\frac{x}{(1+x^2)^2}\varphi(\epsilon x)
=\frac{1}{2} \varphi'(0),
$$
where the second equality is achieved by rescaling $x$.
This shows that 
$$
P_\epsilon(\frac{1}{x})\delta_\epsilon(x)\approx -\frac{1}{2}\delta'(x)
\footnote{It can be shown that the result holds in general for an 
arbitrary embedding}.
$$
As we will see in the next section  this is precisely what happens in the
case of the extended Kerr-geometry namely the delta-function contribution from
the background Ricci tensor combines with the principal value of the
profile to produce derivatives of the delta-function.

\section*{\large\bit 3) Energy-momentum tensor of the maximally analytic
Kerr geometry}

As already pointed out in the introduction, the flat background in the 
Kerr-Schild decomposition of the Kerr-geometry becomes singular under 
the extension to the maximally analytic manifold. The strategy we are going to 
propose in this chapter is to regularize (embed into $\mathcal G$) the 
Kerr-geometry such that it stays in the generalized Kerr-Schild class. \par
Using the same method as in the two-dimensional example presented in 
section one, namely replacing $r^2$ by $r^2+\alpha^2$, the background-metric 
and the vector field  become
\begin{align}\label{regback}
  &ds^2 = - dt^2 +\frac{\Sigma+\alpha^2}{r^2+a^2+\alpha^2}dr^2
+(\Sigma+\alpha^2)d\theta^2 + (r^2+a^2+\alpha^2)\sin^2\theta d\phi^2,
\nonumber\\
&k^a = \partial_t^a + \partial_r^a -\frac{a}{r^2+a^2+\alpha^2}\partial_\phi^a.
\end{align}
Calculating the norm of the latter gives
$$
g_{ab}k^a k^b =-1 + \frac{\Sigma+\alpha^2}{r^2+a^2+\alpha^2} 
+ \frac{a^2\sin^2\theta}{r^2+a^2+\alpha^2} = 0,
$$
which is the first condition for belonging to the generalized
Kerr-Schild class. In addition we have to check the geodeticity of $k^a$.
The simplest way to do this is to observe that $(k\nabla)k^a=0$ is
equivalent to $k\rfloor dk=0$ due to the null character of $k^a$.
\begin{align*}
&dk=-\frac{2a^2\cos\theta\sin\theta}{r^2+a^2+\alpha^2}d\theta dr 
- 2 a\cos\theta\sin\theta d\theta d\phi\\
&k\rfloor dk = \frac{2a^2\cos\theta\sin\theta}{r^2+a^2+\alpha^2}d\theta
- \frac{2a^2\cos\theta\sin\theta}{r^2+a^2+\alpha^2}d\theta =0,
\end{align*}
which is the desired result, showing that the deformation still
belongs to generalized Kerr-Schild class. \par
The first step in the evaluation of the Ricci-tensor $\tilde{R}^a{}_b$
(\ref{gksricci}) is the calculation of the background contribution
$R^a{}_b$. Changing the radial coordinate $r$ to $r=a\sinh u$ 
\begin{align}\label{backricci}
&g_{ab}= - dt_a dt_b + \frac{f^2}{g^2} a^2\cosh^2 u du_a du_b +
a^2 f^2 d\theta_a d\theta_b + a^2g^2\sin^2\theta d\phi_a d\phi_b\nonumber\\
&f^2=\sinh^2 u + \cos^2 \theta + \beta^2\quad g^2=\cosh^2 u + \beta^2
\qquad \alpha=a\beta
\end{align}
and using an adapted frame
\begin{align}
&e^t_a= dt_a && E^a_t = \partial^a_t\nonumber\\
&e^u_a=\frac{a f\cosh u}{g}du_a  &&E^a_u=\frac{g}{a f\cosh u}
\partial_u^a\nonumber\\
&e^\theta_a = af d\theta_a &&E^a_\theta=\frac{1}{af}
\partial^a_\theta\nonumber\\
&e^\phi_a = ag\sin\theta d\phi_a &&E^a_\phi = \frac{1}{ag\sin\theta}
\partial^a_\phi
\end{align}
we obtain for the connection and the Riemann tensor
\begin{align}
&\omega^u{}_\theta = -\frac{\cos\theta\sin\theta}{f^2g}du - 
\frac{g \sinh u}{f^2} d\theta && R^u{}_\theta = 
-\frac{\beta^2\cosh u(\sin^2\theta + g^2)}{f^4 g}du d\theta
\nonumber\\
&\omega^u{}_\phi = -\frac{\sinh u\sin\theta}{f}d\phi  
&&R^u{}_\phi = -\frac{\beta^2\cosh u\sin\theta}{f^3}  du d\phi\nonumber\\
&\omega^\theta{}_\phi = -\frac{g\sin\theta}{f}d\phi  
&&R^\theta{}_\phi = -\frac{\beta^2g\sin\theta}{f^3}  d\theta d\phi,
\end{align}
which in turn gives rise to the mixed Ricci density
$$
\sqrt{g}R^a{}_b = \frac{2a\beta^2\cosh u\sin\theta}{f^4}
(g^2\partial^a_u du_b + \sin^2\theta \partial^a_\theta d\theta_b).
$$
Proceeding along the lines of the example presented in the first section
we obtain the associated distribution
$$
\sqrt{g}R^a{}_b = 2\pi a\delta(u)\delta(\cos\theta)
(\partial^a_u du_b + \partial^a_\theta d\theta_b).
$$
After factoring out the $k^a$-dependence the remaining terms
become
\begin{align}\label{mixed}
  \tilde{R}^a{}_b -R^a{}_b &=
  -\frac{1}{2}k^a(2 \hat{f} k^c R_{cb} -\nabla_b h - \hat{f} \Delta k_b 
+ 2\nabla_c \hat{f}  \nabla^ck_b) \nonumber \\
  &+ \frac{1}{2}k_b (\nabla^a h -\nabla^2 \hat{f} k^a + \hat{f} \Delta k^a 
-2\nabla_c \hat{f}  \nabla^ck^a)\nonumber\\
  &+ (h\frac{1}{2}(\nabla_b k^a + \nabla^a k_b) - \hat{f} 
\nabla_c k^a\nabla^c k_b))  + \hat{f} R^a{}_{cdb}k^c k^d,
\end{align}
where
$$
  h:=\hat{f} \nabla k + (k\nabla)\hat{f}\qquad \Delta k_a :=(d*d* + *d*d)k_a 
  = -\nabla^2 k_a + R^b{}_a k_b.
$$
Using the Laplace-Beltrami $\Delta$ operator instead of the covariant Laplacian $\nabla^2$ considerably facilitates the calculation of $\nabla^2 k^a$.
Let us briefly state some useful identities and then present the final result.
\begin{align}
&\nabla_b k^a = \frac{\sinh u(g^2-f^2)}{ag^2 f^2} E_u^a e^u_b +
\frac{\sinh u}{af^2} E_\theta^a e^\theta_b + \frac{\sinh u}{ag^2}
E_\phi^a e^\phi_b\nonumber\\
&+ \frac{\sin\theta \cos\theta}{agf^2}(E_\theta^a e^u_b - E_u^a e^\theta_b)
+ \frac{\sinh u\sin\theta}{ag^2f}(E_\phi^a e^u_b + E_u^a e^\phi_b)
+\frac{\cos\theta}{agf}(E_\theta^a e^\phi_b - E_\phi^a e^\theta_b)\nonumber\\
&\nabla^2 \hat{f} = -\frac{8m\beta^2\sinh u}{a^3f^8}(g^2+\sin^2\theta)
\qquad\Delta k_a = \frac{2}{a^2 gf^5}(f^4-2g^2 \beta^2) e^u_a - 
\frac{2\sin\theta }{a^2 gf^2} e^\phi_a \nonumber\\
&R^a{}_{cdb}k^ck^d = \frac{\beta^2}{a^2 f^4}\left( \frac{\sin^2\theta}{g^2}
E_u^a e^u_b + \frac{2\sin^2\theta +g^2}{g^2}E_\theta^a e^\theta_b 
+ \frac{f^2}{g^2}E_\phi^a e^\phi_b + \right.\nonumber\\
&\left.\hspace*{2cm}\frac{f\sin\theta}{g^2}
(E_u^a e^\phi_b + E_\phi^a e^u_b)\right)
\end{align}
Putting everything together gives
\begin{align}\label{regdiff}
\tilde{R}^a{}_b -&R^a{}_b = \frac{2\beta^2}{a^2 f^6} \hat{f}\left[
-(f^2+2\sin^2\theta)\partial_t^a dt_b + 
\frac{3f^2\sin^2\theta + 2\sin^4\theta + f^4}{\sin^2\theta}\right.\nonumber\\
&(\frac{\sin\theta}{g}E_\phi^a)(\frac{\sin\theta}{g}e^\phi_b)
+ (f^2+\sin^2\theta)\partial_t^a(\frac{f}{g}e^u_b) 
- 2(f^2 + \sin^2\theta)\partial_t^a(\frac{\sin\theta}{g}e^\phi_b)\nonumber\\
&\left.+2(f^2+\sin^2\theta)(\frac{\sin\theta}{g}E_\phi^a)dt_b
- (f^2 +\sin^2\theta)(\frac{\sin\theta}{g}E_\phi^a)
(\frac{\sin\theta}{g}e^\phi_b)\right ].
\end{align}
Although the last expression looks pretty complicated only 
a small number of terms survive the limiting process 
$\beta\to 0$. The final form of the Ricci-density is
\begin{align}\label{ricfinal}
  \sqrt{\tilde{g}}\tilde{R}^a{}_b &=\nonumber\\ 
&2\pi\delta(\cos\theta)\left(
 a\delta(u)(\partial^a_u du_b + \partial^a_\theta d\theta_b) + m\delta'(u)
 (\partial^a_t -\frac{1}{a}\partial^a_\phi)
(dt_b + ad\phi_b)\right).
\end{align}
The typical integrals one has to deal with in the evaluation of (\ref{regdiff})
are 
\begin{align*}
&\lim\limits_{\beta\to 0}\int \frac{\beta^2\sinh u}{f^6} 
\>\bar{\varphi}(u,\theta)dud\theta =
\int\limits_{-\infty}^\infty du\int\limits_{-\frac{\pi}{2}}^
{\frac{\pi}{2}}d\lambda \frac{\beta^2\sinh u}{f^6}\>\bar{\varphi}(u,
\frac{\pi}{2}-\lambda) \\
&=\int\limits_{-\infty}^\infty du\int\limits_{-\infty}^\infty d\lambda
\frac{u}{\beta(u^2+\lambda^2+1)^3}
\bar{\varphi}(\beta u,\frac{\pi}{2}-\beta \lambda)
=\frac{\pi}{4}\partial_u\bar{\varphi}(0,\frac{\pi}{2})\\
\intertext{which implies}
&\lim\limits_{\beta\to 0}\frac{\beta^2\sinh u}{f^6} = 
-\frac{\pi}{4}\delta'(u)\delta(\cos\theta)
\end{align*}

Several comments are in order. First of all one might wonder
why we derived the Ricci-density instead of the Ricci-tensor
as in \cite{BaNa2}. A simple answer is that the latter has no
associated distribution. However, since only the
determinant of the background metric enters in $\sqrt{\tilde{g}}$ 
our result actually does not differ in this regard from \cite{BaNa2}.
It merely is a question of whether one considers the delta-``function'' to
be a scalar or a density. Our result is also very similar
to that obtained in \cite{Isr2}, in that  the delta-prime term
of (\ref{ricfinal}) becomes negative in the negative $r$-region.
Moreover the tensor-structure of the $m$-dependent terms coincides
with those in \cite{Isr2}. The main difference is the presence of
the background-contribution, which reflects the fact that
the background itself is non-flat, thereby once again emphasizing the fact 
that the extended Kerr-geometry belongs to the generalized Kerr-Schild class 
rather than the Kerr-Schild class. Due to its tensor structure 
the background does not contribute to the mass or angular momentum 
obtained by hooking the corresponding Killing vector into (\ref{ricfinal}).

\section*{Conclusion}

In this work we made use of the generalized Kerr-Schild class to calculate
the energy-momentum tensor for the extension of the Kerr-geometry that
contains the negative mass region. It might seem somewhat surprising to
use the generalized Kerr-Schild class, since Kerr is usually considered to
be a member of the (normal) Kerr-Schild class. However, a closer 
(distributional) look at the ``flat'' part of the decomposition 
reveals that it develops a branch-singularity upon extending to the 
negative $r$-sheet, and that it is therefore no longer flat. For the 
same reason we may no longer consider it as part of the smooth structure.
Using techniques developed for the multiplication of singular
expressions (Colombeau's algebra of new generalized functions) we
show that it is possible obtain a distributional energy-momentum
tensor for the extended geometry, which is concentrated on the
ring-singularity. Our result is consistent with \cite{Isr2}. The main
difference arises from the presence of the background curvature
terms. Due to their tensor structure the latter do not contribute to 
the momentum and and angular momentum densities. A natural further 
line of investigation would be the construction of the ultrarelativistic
limits. However, since the background is no longer flat, the concept
of boosts as its isometries has to be reconsidered.

\vfill
\noindent
{\bf Acknowledgement:} The author wants to thank Werner Israel for his
en\-courage\-ment and numerous sti\-mulating discussions and the 
Depart\-ment of Physics and Astronomy at the University of Victoria for the
hospitality during the final stages of this work.

\end{document}